\documentclass[11pt,a4paper]{article}
\pdfoutput=1
\usepackage[utf8]{inputenc}
\usepackage{jheppub}
\hypersetup{unicode=true,bookmarksopen=true}

\usepackage{amsthm}
\usepackage{mathtools}
\usepackage{lmodern}
\usepackage[T1]{fontenc}
\usepackage{bbm}
\DeclareMathAlphabet{\mathbfi}{OML}{cmm}{b}{it}

\usepackage[nointegrals]{wasysym}

\usepackage{xcolor}

\let\originalleft\left
\let\originalright\right
\renewcommand{\left}{\mathopen{}\mathclose\bgroup\originalleft}
\renewcommand{\right}{\aftergroup\egroup\originalright}
\makeatletter
\newcommand{\biggg}{\bBigg@\thr@@}
\newcommand{\Biggg}{\bBigg@{3.5}}
\makeatother

\makeatletter
\newenvironment{equations}[1][]{\subequations\ifx\relax#1\relax\else\label{#1}\fi\align\ignorespaces}{\endalign\ignorespacesafterend\endsubequations}
\def\@spliteq#1{\begin{equation}\begin{split}#1\end{split}\end{equation}}
\def\@spliteqstar#1{\begin{equation*}\begin{split}#1\end{split}\end{equation*}}
\def\splitequation{\collect@body\@spliteq}
\expandafter\def\csname splitequation*\endcsname{\collect@body\@spliteqstar}

\expandafter\def\csname endsplitequation*\endcsname{\ignorespacesafterend}
\makeatother

\renewcommand{\vec}[1]{{\ifnum9<1#1\mathbf{#1}\else\ifcat\noexpand#1\relax\boldsymbol{#1}\else\mathbfi{#1}\fi\fi}}
\newcommand{\mathe}{\mathrm{e}}
\newcommand{\mathi}{\mathrm{i}}
\let\oldre\Re
\let\oldim\Im
\renewcommand{\Re}{\oldre\mathfrak{e}\,}
\renewcommand{\Im}{\oldim\mathfrak{m}\,}
\newcommand{\total}{\mathop{}\!\mathrm{d}}
\newcommand{\abs}[1]{{\left\lvert{#1}\right\rvert}}
\newcommand{\norm}[1]{{\left\lVert{#1}\right\rVert}}
\newcommand{\arcoth}{\operatorname{arcoth}}
\newcommand{\1}{\mathbbm{1}}
\newcommand{\tr}{\operatorname{tr}}
\newcommand{\eqend}[1]{\,#1}
\newcommand{\bra}[1]{\left\langle{#1}\right\vert}
\newcommand{\ket}[1]{\left\vert{#1}\right\rangle}
\DeclareMathOperator{\im}{im}

\allowdisplaybreaks

\makeatletter
\gdef\@fpheader{\strut}
\makeatother

\begin{document}

\title{Relating the modular Hamiltonian to two-point functions}

\author{Markus B. Fröb}

\affiliation{Institut f{\"u}r Theoretische Physik, Universit{\"a}t Leipzig, Br{\"u}derstra{\ss}e 16, 04103 Leipzig, Germany}

\emailAdd{mfroeb@itp.uni-leipzig.de}

\abstract{We consider the modular Hamiltonian associated to standard subspaces for a free scalar field in a globally hyperbolic spacetime in an arbitrary Gaussian state. We show how the modular Hamiltonian is related to the two-point function of the theory. For the restriction of the modular Hamiltonian to the subspace, we recover formulas that were obtained previously by Peschel, Casini and Huerta. We also show how the same results can be obtained more directly from the KMS condition, and generalize our results to general CCR algebras.}


\maketitle

\section{Introduction}
\label{sec:intro}

The Tomita--Takesaki theory of modular flows of von Neumann algebras~\cite{tomita1967,takesaki1970} has found a multitude of applications in quantum field theory. In particular, it allows to generalize certain concepts and quantities that are defined in terms of density matrices and traces for finite-dimensional quantum systems to relativistic field theory, where they would be otherwise ill-defined without imposing an ultraviolet cutoff. A prominent example is relative entropy, which given a Hilbert space $\mathcal{H}$ and two density matrices $\rho$ and $\sigma$ is defined as
\begin{equation}
\label{eq:srel_def}
S_\text{rel}\left( \rho \Vert \sigma \right) = \tr_\mathcal{H}\left( \rho \ln \rho - \rho \ln \sigma \right) \eqend{,}
\end{equation}
and which is a measure for the amount of information needed to distinguish two states. More precisely, given a system in a state described by $\rho$, the probability of mistakenly ascribing to it the state described by $\sigma$ after a large number $N$ of measurements has been performed is proportional to $\exp\left[ - N S_\text{rel}\left( \rho \Vert \sigma \right) \right]$.

While each of the ingredients in Eq.~\eqref{eq:srel_def} is on its own ill-defined in a quantum field theory (or more specifically for a theory whose local von Neumann algebras are of type III), it is possible to rewrite the equation in such a way that it remains well-defined also in the latter case~\cite{araki1975,araki1976,uhlmann1977}. Namely, one considers instead of $\mathcal{H}$ the bipartite Hilbert space $\mathcal{H} \otimes \mathcal{H}$, and derives the density matrices $\rho$ and $\sigma$ as partial traces of two cyclic and separating states $\ket{\Phi}, \ket{\Psi} \in \mathcal{H} \otimes \mathcal{H}$ (such states can be constructed by purification). The relative entropy~\eqref{eq:srel_def} can then be written as
\begin{equation}
S_\text{rel}\left( \rho \Vert \sigma \right) = - \tr\left( \rho \ln \Delta_{\rho\vert\sigma} \right) = - \omega_\rho\left( \ln \Delta_{\rho\vert\sigma} \right) \eqend{,}
\end{equation}
where
\begin{equation}
\ln \Delta_{\rho\vert\sigma} = \ln \sigma \otimes \1 - \1 \otimes \ln \rho
\end{equation}
is the so-called relative modular Hamiltonian. Tomita--Takesaki theory shows that this operator is well-defined also in quantum field theory, and thus the definition
\begin{equation}
S_\text{rel}\left( \Phi \Vert \Psi \right) = - \omega_\Phi\left( \ln \Delta_{\Phi\vert\Psi} \right) = - \bra{\Phi} \ln \Delta_{\Phi\vert\Psi} \ket{\Phi}
\end{equation}
makes sense for two arbitrary cyclic and separating states $\ket{\Phi}$ and $\ket{\Psi}$ in field theory.

It is not our task nor our goal to give a lengthy introduction to Tomita--Takesaki modular theory and the many beautiful identities and properties that the modular operators have. Instead, we focus on the part that seems most important for practical applications, namely the actual computation of $\ln \Delta_{\Phi\vert\Psi}$. An important case for applications is if the states $\ket{\Phi}$ and $\ket{\Psi}$ are obtained as unitary excitations of a common state $\ket{\Omega}$, in which case the relative modular operator $\ln \Delta_{\Phi\vert\Psi}$ can be obtained by a unitary conjugation of the modular operator $\ln \Delta_\Omega$ of the common state. This in turn is known in a variety of cases, with the most famous one probably being for the algebra of fields restricted to a wedge in Minkowski spacetime and the Minkowski vacuum state (the Bisognano--Wichmann theorem~\cite{bisognanowichmann1975,bisognanowichmann1976}). While this result holds for arbitrary Wightman theories, other examples only apply to free theories or in lower dimensions. In even dimensions, we mention massless free scalar fields in the Minkowski vacuum state and the algebra of fields inside the future light cone~\cite{buchholz1977} or a double cone~\cite{hisloplongo1982,hislop1988}, and in general dimensions scalar fields in the de~Sitter Euclidean (or Bunch--Davies) vacuum state and the algebra of fields inside the static patch of de~Sitter spacetime (the de~Sitter wedge)~\cite{brosepsteinmoschella1998,dreyer1996,florig1999,borchersbuchholz1999,buchholzdreyerflorigsummers2000,figarihoeghkrohnnappi1975,gibbonshawking1977,chandrasekaranlongopeningtonwitten2023}, as well as the analogous result for anti-de Sitter spacetime~\cite{buchholzflorigsummers2000} or FLRW spacetimes~\cite{buchholzmundsummers2001,buchholzmundsummers2002}. For the Schwarzschild black hole, the modular Hamiltonian is known for free scalar fields in the Hartle--Hawking state of the Kruskal extension of the Schwarzschild metric, with the algebra of fields localized in either the right or left Kruskal wedge~\cite{sewell1982,kay1985a,kay1985b,kaywald1991}. The results for massless free fields restricted to either the future light cone or a double cone in Minkowski spacetime can also be generalized to conformal fields in conformally flat spacetimes~\cite{froeb2023}, and many more examples are known in one or two dimensions, both for free fields (scalars and fermions) and conformal field theories in various quantum states and for the algebras of fields restricted to complicated subregions (also including defects and boundaries), see for example Refs.~\cite{borchersyngvason1999,peschel2003,casinihuerta2008,casinihuerta2009a,casinihuerta2009b,longomartinettirehren2010,casinihuerta2011,rehrentedesco2013,cardytonni2016,ariasblancocasinihuerta2017,casinitestetorroba2017,tonnirodriguezlagunasierra2018,ariascasinihuertapontello2018,abterdmenger2018,eislertonnipeschel2019,hollands2021,blancopereznadal2019,friesreyes2019a,friesreyes2019b,digiuliotonni2020,eislerdigiuliotonnipeschel2020,erdmengerfriesreyessimon2020,mintchevtonni2021a,mintchevtonni2021b,gravakelstonni2021,javerzattonni2022,eislertonnipeschel2022,mintchevtonni2022,abateblancokoifmanpereznadal2023,rottolimurcianotonnicalabrese2023,digiulioerdmenger2023,huertavandervelde2023a,huertavandervelde2023b} and references therein.

Let us also mention a technical point: while (smeared) fermion fields are bounded operators, bosonic fields are unbounded even after smearing. To obtain a von Neumann algebra of bounded operators and apply the tools of Tomita--Takesaki theory, one thus has to consider exponentiated fields, the Weyl operators $W(f)$ which satisfy the product rule
\begin{equation}
W(f) W(g) = \mathe^{- \mathi \sigma(f,g)} W(f+g)
\end{equation}
for a symplectic form $\sigma$. In a Fock representation, these are obtained as $\mathe^{\mathi \phi(f)}$ where $\phi$ is the representation of the usual scalar field, and the product rule substitutes for the canonical commutation relations. If $f$ is real (and $\phi$ a Hermitean field), they are unitary operators and thus bounded, but if $f$ is permitted to be complex even the Weyl operators are unbounded. We thus have to restrict to real $f$ and Hermitean $\phi$, but since we can decompose a complex bosonic field (or a complex $f$) into two independent real parts, this is fortunately no restriction.

In this work, we consider a free scalar field theory in a general curved spacetime. We work with the formulation in terms of initial-value data on a Cauchy surface $\Sigma$, and the von Neumann algebra that we consider is generated by Weyl operators $W(f)$ with $f$ restricted to some region $R \subset \Sigma$. For free theories, everything descends to the one-particle Hilbert space $\mathcal{H}$ and the subspace $\mathcal{L}$ corresponding to initial-value data supported in $R$. We show how one can relate the abstract expression for the modular Hamiltonian $\ln \Delta$ in terms of the projector $P$ on the standard subspace $\mathcal{L}$ elaborated in detail in Refs.~\cite{rieffelvandaele1977,ciollilongoruzzi2019,bostelmanncadamurodelvecchio2022,longo2022,bostelmanncadamurominz2023}, namely~\eqref{eq:lndelta_arcoth}
\begin{equation}
\label{eq:intro_lndelta_arcoth}
\ln \Delta = 2 \arcoth\left( \1 - P + I P I \right) \eqend{,}
\end{equation}
to a concrete formula for the integral kernel of its restriction to the subspace in terms of the two-point function $G$. This formula reads~\eqref{eq:ilndelta_subspace_mn}
\begin{equation}
\label{eq:intro_ilndelta_subspace_mn}
I \ln \Delta \bigr\rvert_R = \begin{pmatrix} 0 & 2 M \\ - 2 N & 0 \end{pmatrix}
\end{equation}
with~\eqref{eq:lndelta_mn_def}
\begin{equation}
M = \Pi^\frac{1}{2} B^{-1} \arcoth(2 B) \Pi^\frac{1}{2} \eqend{,} \quad N = \Pi^{- \frac{1}{2}} B \arcoth(2 B) \Pi^{- \frac{1}{2}} \eqend{,}
\end{equation}
where $B = \sqrt{ \Pi^\frac{1}{2} X \Pi^\frac{1}{2} }$ and $X$ and $\Pi$ are the correlation functions of the field and its conjugate momentum in the region $R$, seen as convolution operators. A very similar formula is known from the work of Casini and Huerta~\cite{casinihuerta2009b}, namely
\begin{equation}
M = \Pi \frac{1}{2 C} \ln\left( \frac{2 C + \1}{2 C - \1} \right) \eqend{,} \quad N = \frac{1}{2 C} \ln\left( \frac{2 C + \1}{2 C - \1} \right) X \eqend{,}
\end{equation}
where $C = \sqrt{ X \Pi }$. For the discretized theory where $X$ and $\Pi$ are bounded operators and which is treated in Ref.~\cite{casinihuerta2009b}, we show that both are equivalent. For fermions, a similar connection was shown very recently~\cite{cadamurofroebminz2023}, but for bosons such a proof is still missing. We also generalize our results to general CCR algebras, and finally show that the main relations can alternatively be obtained from the KMS condition that the state $\omega$ satisfies with respect to the modular flow.

While the formula~\eqref{eq:intro_lndelta_arcoth} can be used for numerical computations~\cite{bostelmanncadamurominz2023}, to determine the modular Hamiltonian analytically it seems much more expedient to use the relation~\eqref{eq:intro_ilndelta_subspace_mn}. Indeed, using the spectral decomposition of the two-point function (seen as a convolution operator) and functional calculus, Casini and Huerta~\cite{casinihuerta2009a} computed the explicit form of the modular Hamiltonian for massless fermions in 1+1 dimensions and the Minkowski vacuum state, in the case where the subspace $\mathcal{L}$ corresponds to initial data with support in a collection of intervals. Later on, the same method was used to compute corrections for small mass~\cite{ariasblancocasinihuerta2017,cadamurofroebminz2023}, and similarly the analogue of Eq.~\eqref{eq:intro_ilndelta_subspace_mn} for the $U(1)$ current on the line was used to compute the modular Hamiltonian for various intervals~\cite{ariascasinihuertapontello2018}. Instead of performing the spectral decomposition explicitly, it is also possible to define the logarithm as an integral over the resolvent (our result~\eqref{eq:ilndelta_resolvent_integral}), which seems to be a very efficient method of obtaining the modular Hamiltonian in a variety of settings, see for example the recent works~\cite{blancogarbarzpereznadal2019,erdmengerfriesreyessimon2020,cadamurofroebpereznadal2024} and references therein that treat fermions. Moreover, it is possible to consider a finite-dimensional approximation of the right-hand side (e.g., by applying a lattice discretization), determining the corresponding modular Hamiltonian and finally taking the limit where the approximation becomes exact. Results using this approach can be found in Refs.~\cite{javerzattonni2022,eislertonnipeschel2022,rottolimurcianotonnicalabrese2023,katsinispastras2024} and references therein.

\section{Free scalar fields}

We consider a globally hyperbolic spacetime with Cauchy surface $\Sigma$. The quantization of a free scalar field on such a spacetime is well-known (see for example Ref.~\cite{kaywald1991} and references therein), and we only recapitulate the main features. Initial data for the Klein--Gordon equation is given by a pair of real-valued, smooth and compactly supported functions $f = (f_1,f_2) \in S = C^\infty_{0,\mathbb{R}}(\Sigma) \oplus C^\infty_{0,\mathbb{R}}(\Sigma)$, and the classical time evolution preserves the (non-degenerate) symplectic form
\begin{equation}
\label{eq:symplectic_form}
\sigma(f,g) = \frac{1}{2} \int_\Sigma \left( f_1 g_2 - g_1 f_2 \right) \total x \eqend{.}
\end{equation}
The quantization of this system proceeds as follows~\cite[Prop.~3.1]{kaywald1991}: consider a bilinear positive symmetric form $\mu$ on $S \times S$ satisfying the inequality
\begin{equation}
\label{eq:sigma_mu_inequality}
\sigma^2(f,g) \leq \mu(f,f) \mu(g,g)
\end{equation}
for all $f,g \in S$. The inequality is strict if the state is mixed, but for pure states equality might be obtained. Since a mixed state can always be represented as a pure state on a larger Hilbert space (so-called purification)~\cite[Ch.~4]{petz1990}, we restrict in the following to pure states. Then there exists a complex Hilbert space $\mathcal{H}$ with complex structure $I$ (a bounded operator satisfying $I^2 = - \1$ and $I^\dagger = - I$ with respect to $\mu$) such that $S$ is dense in $\mathcal{H}$ and that
\begin{equation}
\label{eq:scalar_product_sigma_mu}
\left\langle f, g \right\rangle = \mu(f,g) + \mathi \sigma(f,g)
\end{equation}
for all $f,g \in S$. Complex multiplication in $\mathcal{H}$ is defined by $\mathi f = I f$, which entails in particular that $\mu(f, g) = \sigma(f, I g)$. The full Hilbert space of the theory is then constructed as the usual symmetric Fock space over this one-particle Hilbert space, but we only need the one-particle Hilbert space $\mathcal{H}$ in the following. The symmetry of $\mu$ and the antisymmetry of $\sigma$ also entail that
\begin{equation}
\label{eq:complex_structure_sigma_mu}
\sigma(f, g) = - \mu(f,I g) = - \mu(I g,f) = - \sigma(I g, I f) = \sigma(I f, I g) \eqend{,} \quad \mu(I f, I g) = \mu(f, g) \eqend{.}
\end{equation}
The relations~\eqref{eq:complex_structure_sigma_mu} easily yield that $I^\dagger = - I$, where the adjoint is computed either with respect to the full complex scalar product or its real part $\mu$. We note that the construction of a state (and thus a complex structure) in a general spacetime is a non-trivial task, see for example the recent Refs.~\cite{gerardwrochna2020,muchoeckl2022,sanchezsanchezschrohe2023,islamstrohmaier2024,derezinskigass2024} and the many references therein.

Consider now a closed real-linear subspace $\mathcal{L} \subset \mathcal{H}$. We assume that $\mathcal{L}$ is standard and factorial, which means that $\mathcal{L} \cap I \mathcal{L} = \{ 0 \}$ (separating), that $\mathcal{L} + I \mathcal{L}$ is dense in $\mathcal{H}$ (cylic), such that $\mathcal{H} = \mathcal{L} \oplus I \mathcal{L}$, and that $\mathcal{L} \cap (I \mathcal{L})^\perp = \{ 0 \}$, where the orthogonal complement is taken with respect to the real part $\mu$ of the scalar product. It follows straightforwardly that also $\mathcal{L}^\perp \cap I \mathcal{L} = \{ 0 \}$ and $\mathcal{L}^\perp \cap (I \mathcal{L})^\perp = \mathcal{L}^\perp \cap I \mathcal{L}^\perp = \{ 0 \}$\footnote{For the first equality, consider $g \in (I \mathcal{L})^\perp \Leftrightarrow \mu(g, I f) = 0 \,\forall\, f \in \mathcal{L} \Leftrightarrow \mu(I g, f) = 0 \,\forall\, f \in \mathcal{L} \Leftrightarrow I g \in \mathcal{L}^\perp \Leftrightarrow g \in I \mathcal{L}^\perp$, using that $I^2 = - \1$ and $\mu(I f, I g) = \mu(f,g)$.}. Namely, for $f \in \mathcal{L}^\perp \cap I \mathcal{L}$ we have $f = I h$ for some $h \in \mathcal{L}$ and $\mu(f,g) = 0 \,\forall\, g \in \mathcal{L}$, hence $\mu(h, I g) = 0 \,\forall\, g \in \mathcal{L}$ using that $I^\dagger = -I$ and thus $h \in \mathcal{L} \cap (I \mathcal{L})^\perp = \{ 0 \}$ and $f = I 0 = 0$. The second relation is shown analogously, and the four subspaces $\mathcal{L}$, $I \mathcal{L}$, $\mathcal{L}^\perp$ and $(I \mathcal{L})^\perp$ are thus in generic position~\cite{dixmier1948,halmos1969}. In applications, such a subspace corresponds to functions with support inside a region $R \subset \Sigma$, for example a half-space (the initial data for a wedge) or a sphere (the initial data for a double cone). In order for $\mathcal{L}$ to be cylic, $I$ must then be an anti-local operator~\cite{segalgoodman1965,masuda1972,murata1973}, which in particular entails that $I f$ has support in all of $\Sigma$ even if $f$ is supported only in $R$. Let $P$ be the operator that multiplies $h \in \mathcal{H}$ with the characteristic function of $R$ such that $\im P = \mathcal{L}$. We compute
\begin{equation}
\mu\left( (1-P) f, I P g \right) = - \sigma\left( (1-P) f, P g \right) = 0 \eqend{,}
\end{equation}
where the last equality holds because of the explicit expression~\eqref{eq:symplectic_form} of the symplectic form. It follows that $\im (1-P) = \ker P = (I \mathcal{L})^\perp$, and thus $P$ is not an orthogonal but a symplectic projection (called ``cutting projection'' in~\cite{longo2022}). Since $\mathcal{H} = \mathcal{L} \oplus I \mathcal{L}$, any vector $h \in \mathcal{H}$ can be uniquely written as $h = f + I g$ with $f,g \in \mathcal{L}$. Using the symplectic projection $P$, we also obtain the decomposition $h = f + g$ with $f = P h \in \mathcal{L}$ and $g = (1-P) h \in (I \mathcal{L})^\perp$, such that $\mathcal{H} = \mathcal{L} \oplus (I \mathcal{L})^\perp$ also holds. Multiplying by the complex structure, we further obtain the decomposition $h = f + g$ with $f = - I P I h \in I \mathcal{L}$ and $g = (1 + I P I) h$, and since $\mu(f, (1 + I P I) h) = \mu(I f, (1-P) I h) = 0 \,\forall\, f \in \mathcal{L}$ because of $\im (1-P) = (I \mathcal{L})^\perp$, we have $g = (1 + I P I) h \in \mathcal{L}^\perp$ and it holds that $\mathcal{H} = \mathcal{L}^\perp \oplus I \mathcal{L}$. To determine the adjoint $P^\dagger$ of the symplectic projection with respect to $\mu$, we take $f + g \in \mathcal{L} \oplus (I \mathcal{L})^\perp$ and $h + k \in \mathcal{L}^\perp \oplus I \mathcal{L}$, and compute
\begin{splitequation}
&\mu\left( P^\dagger (h+k), f+g \right) + \mu\left( I P I (h+k), f+g \right) \\
&= \mu\left( h+k, P (f+g) \right) - \mu\left( P I (h + k), I (f+g) \right) \\
&= \mu\left( h+k, f \right) - \mu\left( I k, I (f+g) \right) = 0 \eqend{,}
\end{splitequation}
where we used that $\mu(I f, I g) = \mu(f,g)$ and that $I h \in (I \mathcal{L})^\perp$ such that $P I h = 0$. It follows that $P^\dagger = - I P I$, and clearly $P$ is not an orthogonal projection. Moreover, with the above results we have $\im P^\dagger = I \mathcal{L}$ and $\ker P^\dagger = \mathcal{L}^\perp$.

\subsection{Modular data}

For a free scalar field in the Fock representation, the modular data is obtained as the second quantization of the one-particle modular data~\cite{figlioliniguido1989,figlioliniguido1994}, such that we can restrict to the one-particle level. In the above standard subspace setting, all the modular objects have been determined and their properties elaborated~\cite{rieffelvandaele1977,ciollilongoruzzi2019,bostelmanncadamurodelvecchio2022,longo2022,bostelmanncadamurominz2023}, and we will give just a summary of the main formulas and derivations.

Using that any vector $h \in \mathcal{H}$ can be uniquely written as $h = f + I g$ with $f,g \in \mathcal{L}$, the Tomita operator $S$ is well-defined as the closure of the densely defined anti-linear involution
\begin{equation}
S \colon f + I g \mapsto f - I g \eqend{.}
\end{equation}
Its polar decomposition results in the modular conjugation $J$ and the modular operator $\Delta$, and we would like to obtain a formula for $\Delta$ in terms of the projector $P$ on the standard subspace $\mathcal{L}$. For this, we write $\Delta = S^\dagger S$ and consider
\begin{splitequation}
\left\langle S (f + I g), h + I k \right\rangle &= \left\langle f - I g, h + I k \right\rangle = \mu\left( f - I g, h + I k \right) + \mathi \sigma\left( f - I g, h + I k \right) \\
&= \mu\left( h + I k, f - I g \right) + \mathi \mu\left( h + I k, I f + g \right)
\end{splitequation}
for $h + I k \in ( I \mathcal{L} )^\perp \oplus \mathcal{L}^\perp$ and $f + I g \in \mathcal{L} \oplus I \mathcal{L}$, where we used the symmetry of $\mu$ and that $\sigma(f,g) = \mu(f,-I g)$. Using that $h,k \in ( I \mathcal{L} )^\perp$ and $f,g \in \mathcal{L}$ as well as $\mu(f,g) = \mu(I f, I g)$, this reduces to
\begin{splitequation}
\left\langle S (f + I g), h + I k \right\rangle &= \mu\left( h, f \right) - \mu\left( k, g \right) + \mathi \mu\left( h, g \right) + \mathi \mu\left( k, f \right) \\
&= \mu\left( h - I k, f + I g \right) + \mathi \mu\left( h - I k, g - I f \right) \\
&= \mu\left( h - I k, f + I g \right) + \mathi \sigma\left( h - I k, I g + f \right) = \left\langle h - I k, f + I g \right\rangle \eqend{.}
\end{splitequation}
On the other hand, for an antilinear operator we have
\begin{equation}
\left\langle S f, g \right\rangle = \left\langle f, S^\dagger g \right\rangle^* = \left\langle S^\dagger g, f \right\rangle
\end{equation}
for all $f,g \in \mathcal{H}$, and thus can identify the action
\begin{equation}
S^\dagger (h + I k) = h - I k \eqend{,}
\end{equation}
recalling that $h + I k \in ( I \mathcal{L} )^\perp \oplus \mathcal{L}^\perp$. 

To derive the relation between $\Delta$ and $P$, we first would like to show that~\cite{ciollilongoruzzi2019}
\begin{equation}
P \left( \1 - \Delta \right) = \1 + S \eqend{.}
\end{equation}
Using that for $f + I g \in \mathcal{L} \oplus I \mathcal{L}$ we have
\begin{splitequation}
\left[ P \left( \1 - \Delta \right) - \1 - S \right] ( f + I g ) &= P ( f + I g ) - P S^\dagger ( f - I g ) - 2 f \\
&= - P ( \1 + S^\dagger ) ( f - I g ) \eqend{,}
\end{splitequation}
we compute for $h \in \mathcal{H}$ that
\begin{splitequation}
&\left\langle h , \left[ P \left( \1 - \Delta \right) - \1 - S \right] ( f + I g ) \right\rangle \\
&= - \left\langle h, P ( \1 + S^\dagger ) ( f - I g ) \right\rangle \\
&= - \mu\left( h, P ( \1 + S^\dagger ) ( f - I g ) \right) - \mathi \mu\left( I h, P ( \1 + S^\dagger ) ( f - I g ) \right) \\
&= \mu\left( I P I h, ( \1 + S^\dagger ) ( f - I g ) \right) - \mathi \mu\left( I P h, ( \1 + S^\dagger ) ( f - I g ) \right) \eqend{,}
\end{splitequation}
where we used that $\mu\left( f, P g \right) = \mu\left( P I f, I g \right)$ and that $\mu(I f, I g) = \mu(f,g)$. Using furthermore that for the real part of the scalar product we have
\begin{equation}
\mu\left( f, S^\dagger g \right) = \mu\left( S^\dagger g, f \right) = \mu\left( S f, g \right) = \mu\left( g, S f \right)
\end{equation}
and that $S I h = - I S h$ for all $h \in \mathcal{H}$, it follows that
\begin{splitequation}
&\left\langle h, \left[ P \left( \1 - \Delta \right) - \1 - S \right] ( f + I g ) \right\rangle \\
&= \mu\left( ( \1 + S ) I P I h, f - I g \right) - \mathi \mu\left( ( \1 + S ) I P h, f - I g \right) \\
&= \mu\left( I ( \1 - S ) P I h, f - I g \right) - \mathi \mu\left( I ( \1 - S ) P h, f - I g \right) = 0 \eqend{,}
\end{splitequation}
where the last equality holds because $P f \in \mathcal{L}$ such that $S P f = P f$.

The analogous computation shows that
\begin{equation}
I P I \left( \1 - \Delta \right) = - \1 + S \eqend{,}
\end{equation}
and combining both we obtain
\begin{equation}
\label{eq:relation_projector_delta}
\1 - P + I P I = - \left( \1 + \Delta \right) \left( \1 - \Delta \right)^{-1} \eqend{,}
\end{equation}
from which it follows that~\cite{bostelmanncadamurominz2023}
\begin{equation}
\label{eq:lndelta_arcoth}
\ln \Delta = 2 \arcoth\left( \1 - P + I P I \right) \eqend{.}
\end{equation}
This equation needs to be understood from spectral calculus. In particular, since $\Delta$ is a positive self-adjoint operator the relation~\eqref{eq:relation_projector_delta} shows that the spectrum of $\1 - P + I P I$ is contained in $(-\infty,-1] \cup [1,\infty)$, such that the $\arcoth$ is well-defined.

\subsection{An alternative formula for the modular Hamiltonian}

While the formula~\eqref{eq:lndelta_arcoth} is short, it is difficult to use in practice, and we would like to derive a different one. For complex numbers, we have the integral representation
\begin{equation}
\label{eq:arccoth_integral}
\arcoth z = \int_1^\infty \frac{z}{t^2 z^2 - 1} \total t \eqend{,}
\end{equation}
which we would like to extend to self-adjoint operators, following~\cite{schmuedgen2012}. Let thus $A$ be a self-adjoint operator on $\mathcal{H}$, whence by the spectral theorem there exists a unique spectral measure $E_A$ on the Borel sigma-algebra of $\mathbb{R}$, supported on the spectrum $\sigma(A)$, such that
\begin{equation}
A = \int_{\sigma(A)} \lambda \total E_A(\lambda) \eqend{.}
\end{equation}
If $\sigma(A) \subseteq (-\infty,-1] \cup [1,\infty)$, spectral calculus defines
\begin{splitequation}
\arcoth(A) &= \int_{\sigma(A)} \arcoth(\lambda) \total E_A(\lambda) \\
&= \int_{\sigma(A)} \int_1^\infty \frac{\lambda}{t^2 \lambda^2 - 1} \total t \total E_A(\lambda) \eqend{.}
\end{splitequation}
For $x \in \mathcal{D}(A)$ and $0 < \epsilon < \frac{1}{2}$, consider the vector
\begin{equation}
y = P_\epsilon x \eqend{,} \quad P_\epsilon = E_A\left( \left[ - \frac{2}{\epsilon}, - 1 - \epsilon \right] \right) + E_A\left( \left[ 1 + \epsilon, \frac{1}{\epsilon} \right] \right) \eqend{.}
\end{equation}
For all such $\epsilon$, $y$ lies in the domain of $A$, and we have
\begin{splitequation}
\arcoth(A) y &= \int_{\sigma(A)} \int_1^\infty \frac{\lambda}{t^2 \lambda^2 - 1} \total t \total E_A(\lambda) y \\
&= \int_1^\infty \int_{\sigma(A)} \frac{\lambda}{t^2 \lambda^2 - 1} \total E_A(\lambda) y \total t \\
&= \int_1^\infty A \left( t^2 A^2 - \1 \right)^{-1} y \total t \eqend{,}
\end{splitequation}
where we could interchange the integrals for $\epsilon > 0$ by Fubini's theorem for Bochner integrals. Since the set of all such $y$ (for all $0 < \epsilon < \frac{1}{2}$) is a core for $A$, the conclusion follows.

Applied to our result~\eqref{eq:lndelta_arcoth}, we thus obtain
\begin{splitequation}
\label{eq:lndelta_integral}
\ln \Delta &= 2 \int_1^\infty \left( \1 - P + I P I \right) \left[ t^2 \left( \1 - P + I P I \right)^2 - \1 \right]^{-1} \total t \\
&= 2 I \int_1^\infty \left( I - I P - P I \right) \left[ \1 + t^2 P I P I P + t^2 (\1-P) I (\1-P) I (\1-P) \right]^{-1} \total t \eqend{.}
\end{splitequation}
We see that $A^2 = P I P I P + ( \1 - P ) I  ( \1 - P ) I ( \1 - P )$ is block diagonal, i.e., that it leaves invariant the closed subspaces $\im P = \mathcal{L}$ and $\im (1-P) = (I \mathcal{L})^\perp$ which together sum to the full Hilbert space. It follows that also the resolvent is block diagonal, such that
\begin{splitequation}
&\left[ \1 + t^2 P I P I P + t^2 ( \1 - P ) I  ( \1 - P ) I ( \1 - P ) \right]^{-1} \\
&= P \left[ \left( \1 + t^2 I P I \right) \Bigr\rvert_\mathcal{L} \right]^{-1} P \\
&\quad+ ( \1 - P ) \left[ \left( \1 + t^2 I ( \1 - P ) I \right) \Bigr\rvert_{(I \mathcal{L})^\perp} \right]^{-1} ( \1 - P ) \eqend{,}
\end{splitequation}
where the inverse is computed on each subspace. Therefore, we obtain
\begin{splitequation}
\label{eq:ilndelta_resolvent_integral}
I \ln \Delta &= 2 \int_1^\infty P I P \left[ \left( \1 + t^2 I P I \right) \Bigr\rvert_\mathcal{L} \right]^{-1} P \total t \\
&\quad- 2 \int_1^\infty ( \1 - P ) I ( \1 - P ) \left[ \left( \1 + t^2 I ( \1 - P ) I \right) \Bigr\rvert_{(I \mathcal{L})^\perp} \right]^{-1} ( \1 - P ) \total t \eqend{.}
\end{splitequation}
We see that $I \ln \Delta$ leaves not only the subspace $\mathcal{L}$ invariant (which is known), but also the real orthogonal complement $(I \mathcal{L})^\perp$.

\subsection{Relation to the two-point function}

To connect with other formulas in the literature which only involve functions supported in the region $R \subset \Sigma$, we need to express the projection of the complex structure $P I P$ using the two-point function. For this, we need to consider a different Hilbert space, namely $\hat{\mathcal{H}} = L^2_\mathbb{C}(R) \oplus L^2_\mathbb{C}(R)$ with the $L^2$ scalar product
\begin{equation}
\left( f, g \right) = \int_R \left( f_1^* g_1 + f_2^* g_2 \right) \total x \eqend{,}
\end{equation}
which we see as the completion of $S_{R,\mathbb{C}} = C^\infty_{0,\mathbb{C}}(R) \oplus C^\infty_{0,\mathbb{C}}(R)$ with respect to this scalar product. Both the symplectic form $\sigma$~\eqref{eq:symplectic_form} and the bilinear form $\mu$ can be straightforwardly extended to real functions $f, g \in S_{R,\mathbb{R}} = C^\infty_{0,\mathbb{R}}(R) \oplus C^\infty_{0,\mathbb{R}}(R)$. Namely, we write
\begin{equation}
\label{eq:symplectic_form_epsilon}
\sigma(f, g) = \frac{1}{2} \left( f, \epsilon g \right)
\end{equation}
with the constant matrix $\epsilon = \begin{pmatrix} 0 & 1 \\ - 1 & 0 \end{pmatrix}$ fulfilling $\epsilon^2 = - \1$ and 
\begin{equation}
\label{eq:bilinear_form_pip}
\mu(f, g) = \sigma(P f, I P g) = \frac{1}{2} \left( f, P \epsilon I P g \right) = \frac{1}{2} \left( f, \epsilon P I P g \right) \eqend{.}
\end{equation}
Since $I$ is an antilocal operator and thus in particular does not preserve the support of functions, we had to introduce the projection $P$, which is the multiplication with the characteristic function of the region $R$ (and thus clearly commutes with $\epsilon$). This is also the reason why we consider the usual complex-valued $L^2$ space instead of simply restricting $I$ to $R$: the restriction does not fulfill anymore the requirements for a complex structure, in particular $(P I P)^2 \neq - \1$. Whereas $P$ is an unbounded operator on the original Hilbert space $\mathcal{H}$, it is the identity on $\hat{\mathcal{H}}$ and thus trivially bounded. The price we have to pay for this simplification is that the projection of the complex structure $P I P$ is unbounded on $\hat{\mathcal{H}}$, while $I$ was bounded on $\mathcal{H}$. However, since $\mu$ is symmetric and positive definite, it follows that $\left( f, \epsilon P I P g \right) = \left( \epsilon P I P f, g \right)$ for real $f$ and $g$. We can thus extend $\epsilon P I P$ to a symmetric, positive and densely defined operator on $S_{R,\mathbb{C}}$, and we obtain a self-adjoint operator on $\hat{\mathcal{H}}$ by taking the Friedrichs extension (denoted by the same symbol). On the other hand, the polarization identity shows that $\epsilon$ cannot be extended as a linear operator to $S_{R,\mathbb{C}}$, but only as an antilinear operator. This is however inconvenient for later use, and so we instead consider $\mathi \epsilon$ as a linear, symmetric and bounded operator on $S_{R,\mathbb{C}}$. One easily verifies that this entails
\begin{equation}
\label{eq:iepsilon_sigma_relation}
\left( f + \mathi g, \mathi \epsilon ( f + \mathi g ) \right) = - 2 \sigma(f,g) \quad\text{for}\quad f, g \in S_{R,\mathbb{R}} \eqend{,}
\end{equation}
from which arbitrary matrix elements can be computed using the polarization identity. Moreover, it is clear that while $\mathi \epsilon$ is symmetric, its matrix elements do not have a definite sign. (In fact, it is easy to see that $\hat{\mathcal{H}}$ decomposes into a direct sum of eigenspaces corresponding to the eigenvalues $\pm 1$ of $\mathi \epsilon$.)

The original scalar product $\langle \cdot,\cdot \rangle$ on $\mathcal{H}$, restricted to $f, g \in S_{R,\mathbb{R}}$, then defines an operator $G$ on $S_{R,\mathbb{R}}$ via
\begin{equation}
\label{eq:g_scalar_product}
\left( f, G g \right) = \left\langle f, g \right\rangle = \frac{1}{2} \left( f, \epsilon P I P g \right) + \frac{1}{2} \left( f, \mathi \epsilon g \right) \eqend{.}
\end{equation}
Using the extensions of $\epsilon P I P$ and $\mathi \epsilon$ as above, we extend $G$ to a symmetric, densely defined and semibounded operator on $S_{R,\mathbb{C}}$, and denote its Friedrichs extension to a self-adjoint operator on $\hat{\mathcal{H}}$ by the same symbol.\footnote{Since the form associated to the bounded operator $\mathi \epsilon$ is closed, and the sum of two closed semibounded forms is closed, the Friedrichs extension of $G$ coincides with the sum of the Friedrichs extension of $\epsilon P I P$ and $\mathi \epsilon$. $G$ is thus a self-adjoint operator on $\hat{\mathcal{H}}$, whose domain of definition is the domain of the Friedrichs extension of $\epsilon P I P$, namely $\mathcal{D}\bigl( ( \epsilon P I P )^\frac{1}{2} \bigr)$.} While we know that $\epsilon P I P$ is positive, $\mathi \epsilon$ can have either sign, and so it is not a priori clear that $G$ is actually a positive operator. Let us show this: by using the relation~\eqref{eq:iepsilon_sigma_relation} and the inequality~\eqref{eq:sigma_mu_inequality}, we compute that
\begin{splitequation}
\left( f + \mathi g, \mathi \epsilon ( f + \mathi g ) \right)^2 &= 4 \sigma(f,g)^2 \leq 4 \mu(f,f) \mu(g,g) \\
&= \left[ \mu(f,f) + \mu(g,g) \right]^2 - \left[ \mu(f,f) - \mu(g,g) \right]^2 \\
&\leq \left[ \mu(f,f) + \mu(g,g) \right]^2 = \left( f + \mathi g, \epsilon P I P ( f + \mathi g ) \right)^2
\end{splitequation}
for $f, g \in S_{R,\mathbb{R}}$, such that
\begin{equation}
\label{eq:epspip_ieps_positive}
\left[ \left( h, \epsilon P I P h \right) - \left( h, \mathi \epsilon h \right) \right] \left[ \left( h, \epsilon P I P h \right) + \left( h, \mathi \epsilon h \right) \right] \geq 0
\end{equation}
for all $h \in \mathcal{D}\bigl( ( \epsilon P I P )^\frac{1}{2} \bigr)$. It follows that $G \geq 0$, and we obtain an isometry
\begin{equation}
U \colon \mathcal{H} \to \hat{\mathcal{H}} \eqend{,} \quad f + I g \mapsto \sqrt{G} f + \mathi \sqrt{G} \, g \quad\text{for}\quad f,g \in \mathcal{L} \eqend{,}
\end{equation}
where we used that $\mathcal{H} = \mathcal{L} \oplus I \mathcal{L}$ such that $U$ is uniquely defined. Rewriting the relation~\eqref{eq:g_scalar_product} as relation between operators on $\hat{\mathcal{H}}$ according to
\begin{equation}
\label{eq:relation_g_i}
2 G = \epsilon P I P + \mathi \epsilon \quad\Leftrightarrow\quad P I P = 2 \mathi ( \mathi \epsilon )^{-1} G - \mathi \1 \eqend{,}
\end{equation}
we may replace the projection of the complex structure $P I P$ in our result~\eqref{eq:ilndelta_resolvent_integral} to obtain
\begin{equation}
\label{eq:ilndelta_resolvent_gi}
U \left( I \ln \Delta \bigr\rvert_R \right) U^{-1} = \mathi K = 2 \mathi \int_1^\infty \frac{2 ( \mathi \epsilon )^{-1} G - \1}{\1 - t^2 \left[ 2 ( \mathi \epsilon )^{-1} G - \1 \right]^2} \total t
\end{equation}
on a suitable dense domain in $\hat{\mathcal{H}}$, which defines the modular Hamiltonian $K$ on $\hat{\mathcal{H}}$. We may interpret both $G$ and $K$ as convolution operators acting on functions with support in $R$, and then the formula~\eqref{eq:ilndelta_resolvent_gi} relates their integral kernels. The analogous computation establishes a similar formula for the restriction of the modular Hamiltonian to the complement region $\Sigma \setminus R$, which only depends on the two-point function $G$ restricted to the complement. We omit the straightforward details, which essentially only differ in an overall minus sign coming from the result~\eqref{eq:ilndelta_resolvent_integral}.

To obtain a more concrete expression for the result~\eqref{eq:ilndelta_resolvent_gi}, we note that of the initial data $f = (f_1,f_2)$ the first component $f_1$ is the initial data for the field $\phi$ itself, while the second component $f_2$ is the initial data for its normal derivative, the conjugate momentum $\pi$. The integral kernel of $G$ is thus nothing else but the two-point function, which decomposes as
\begin{equation}
\label{eq:g_restricted}
G = \begin{pmatrix} X & \frac{\mathi}{2} \1 \\ - \frac{\mathi}{2} \1 & \Pi \end{pmatrix} \eqend{.}
\end{equation}
The off-diagonal entries are one-half of the equal-time commutator between $\phi$ and $\pi$, while the diagonal entries are the equal-time correlation functions of $\phi$ with itself ($X$) or $\pi$ with itself ($\Pi$), restricted to $R$, which are both symmetric. Since the symmetric part of $G$ is equal to the positive operator $\epsilon P I P$, it follows that also $X$ and $\Pi$ are positive operators, and in particular invertible on a suitable domain dense in $\hat{\mathcal{H}}$. From the relation~\eqref{eq:relation_g_i} we further obtain
\begin{equation}
P I P = 2 \mathi ( \mathi \epsilon )^{-1} G - \mathi \1 = \begin{pmatrix} 0 & - 2 \Pi \\ 2 X & 0 \end{pmatrix} \eqend{,}
\end{equation}
and thus the result~\eqref{eq:ilndelta_resolvent_gi} reads
\begin{splitequation}
\label{eq:ilndelta_subspace_mn}
\mathi K &= 4 \begin{pmatrix} 0 & - \Pi \\ X & 0 \end{pmatrix} \int_1^\infty \left[ \1 + 4 t^2 \begin{pmatrix} 0 & - \Pi \\ X & 0 \end{pmatrix}^2 \right]^{-1} \total t \\
&= \begin{pmatrix} 0 & - 4 \Pi^\frac{1}{2} \int_1^\infty \left( \1 - 4 t^2 \Pi^\frac{1}{2} X \Pi^\frac{1}{2} \right)^{-1} \total t \, \Pi^\frac{1}{2} \\ 4 X \Pi^\frac{1}{2} \int_1^\infty \left( \1 - 4 t^2 \Pi^\frac{1}{2} X \Pi^\frac{1}{2} \right)^{-1} \total t \, \Pi^{- \frac{1}{2}} & 0 \end{pmatrix} \\
&= \begin{pmatrix} 0 & 2 M \\ - 2 N & 0 \end{pmatrix}
\raisetag{5em}
\end{splitequation}
with
\begin{equations}[eq:lndelta_mn_def]
M &= \Pi^\frac{1}{2} B^{-1} \arcoth(2 B) \Pi^\frac{1}{2} \eqend{,} \\
N &= \Pi^{- \frac{1}{2}} B \arcoth(2 B) \Pi^{- \frac{1}{2}} \eqend{.}
\end{equations}
Here we defined $B = \sqrt{ \Pi^\frac{1}{2} X \Pi^\frac{1}{2} }$, and employed the integral representation~\eqref{eq:arccoth_integral} of the $\arcoth$, extended to self-adjoint operators. For this, we used that since $\Pi^\frac{1}{2} X \Pi^\frac{1}{2}$ is positive, symmetric and densely defined we can again take its Friedrichs extension, which we denote by the same symbol. To ensure that Eqs.~\eqref{eq:lndelta_mn_def} are valid definitions, we need to verify that $B \geq \frac{1}{2}$, which follows from positivity of the state~\cite{casinihuerta2009b}. To show this, we write the decomposition~\eqref{eq:g_restricted} as
\begin{equation}
G = \begin{pmatrix} \1 & \frac{\mathi}{2} \Pi^{-1} \\ 0 & \1 \end{pmatrix} \begin{pmatrix} \left( X \Pi - \frac{1}{4} \right) \Pi^{-1} & 0 \\ 0 & \Pi \end{pmatrix} \begin{pmatrix} \1 & 0 \\ - \frac{\mathi}{2} \Pi^{-1} & \1 \end{pmatrix} \eqend{,}
\end{equation}
where the inverse $\Pi^{-1}$ is well-defined (at least) on $C_0^\infty(R)$, such that for $f = \left( \Pi^\frac{1}{2} f_1, \frac{\mathi}{2} \Pi^{- \frac{1}{2}} f_1 \right)$ we obtain
\begin{equation}
\left( f, G f \right) = \left( f_1, \left( \Pi^\frac{1}{2} X \Pi^\frac{1}{2} - \frac{1}{4} \right) f_1 \right) \geq 0 \eqend{.}
\end{equation}
Therefore, $\Pi^\frac{1}{2} X \Pi^\frac{1}{2}$ is densely defined and lower semibounded with bound $\frac{1}{4}$, and its Friedrichs extension satisfies the same bound, such that $B \geq \frac{1}{2}$.

The result~\eqref{eq:ilndelta_subspace_mn} together with the explicit expressions~\eqref{eq:lndelta_mn_def} for $M$ and $N$ is very similar to the result of Casini and Huerta~\cite{casinihuerta2009b}, who give the modular Hamiltonian as a second-quantized operator on Fock space. (Similar but less explicit formulas were also obtained by Peschel~\cite{peschel2003} and Araki~\cite{araki1971}.) However, Casini and Huerta work with a discretized theory, where $X$ and $\Pi$ are actually bounded operators. In this case, it is possible to express the operators $M$ and $N$~\eqref{eq:lndelta_mn_def} in a different way. Namely, $X \Pi$ as the product of two positive operators is weakly positive in the sense of Wigner~\cite{wigner1963}, which entails that it is similar to a positive operator. In fact, we see immediately that $\Pi^\frac{1}{2} X \Pi^\frac{1}{2}$ is similar to $X \Pi$ with intertwining operator $\Pi^{- \frac{1}{2}}$, hence both have the same spectrum. It follows that $X \Pi$ has a spectral decomposition of the form~\cite{wigner1963}
\begin{equation}
X \Pi = \int_0^M \lambda \total \left( \Pi^{- \frac{1}{2}} E_\lambda \Pi^\frac{1}{2} \right) \eqend{,}
\end{equation}
where $E_\lambda$ are the self-adjoint projections in the spectral decomposition of $\Pi^\frac{1}{2} X \Pi^\frac{1}{2}$ and $M = \norm{ \Pi^\frac{1}{2} X \Pi^\frac{1}{2} }$. This allows us to define functions of $X \Pi$ by spectral calculus, and we obtain
\begin{equations}[eq:lndelta_mn_bounded]
M &= \Pi C^{-1} \arcoth(2 C) = \Pi \frac{1}{2 C} \ln\left( \frac{2 C + \1}{2 C - \1} \right) \eqend{,} \\
N &= C^{-1} \arcoth(2 C) X = \frac{1}{2 C} \ln\left( \frac{2 C + \1}{2 C - \1} \right) X
\end{equations}
with $C = \sqrt{ X \Pi }$. These now exactly coincide with the result of Casini and Huerta~\cite{casinihuerta2009b}.

\subsection{General CCR algebras}
\label{sec:general_ccr}

Our results can be easily generalized to more general commutation relations, as long as the theory remains free. Relative entropy for general CCR algebras was studied in Refs.~\cite{ariascasinihuertapontello2018,bostelmanncadamurodelvecchio2022}, using both the standard subspace formalism and the generalization of formulas~\eqref{eq:ilndelta_subspace_mn} and~\eqref{eq:lndelta_mn_bounded}. We also refer to~\cite{borchersyngvason1999}, where the modular Hamiltonian of massless generalized free fields in two dimensions has been determined both for thermal states and the vacuum in various regions.

A generalized free field is characterized by a symplectic form $\sigma$ different from~\eqref{eq:symplectic_form}. However, in the standard subspace formulation the concrete form of $\sigma$ is irrelevant, such that the result~\eqref{eq:ilndelta_resolvent_integral} for $I \ln \Delta$ still holds. Moreover, $\sigma$ can still be written in the form~\eqref{eq:symplectic_form_epsilon}, where however now $\epsilon$ is a (possibly unbounded) densely defined and invertible antisymmetric operator on $S_{R,\mathbb{R}}$ (since $\sigma$ is non-degenerate by assumption). We assume that $\epsilon$ and $P$ commute, and again extend $\mathi \epsilon$ to a linear, symmetric and densely defined operator on $\hat{\mathcal{H}}$. Denoting its closure by the same symbol, its polar decomposition reads $\mathi \epsilon = J \abs{ \mathi \epsilon }$ with $J^\dagger = J$ and $J^2 = \1$ since $\epsilon$ is invertible. The relation~\eqref{eq:relation_g_i} between the two-point function $G$, $\mathi \epsilon$ and the projection of the complex structure $P I P$ then still holds, and thus our result~\eqref{eq:ilndelta_resolvent_gi}. Let us define
\begin{splitequation}
\label{eq:u_def}
E &= - \mathi \abs{ \mathi \epsilon }^\frac{1}{2} P I P \abs{ \mathi \epsilon }^{-\frac{1}{2}} = J \abs{ \mathi \epsilon }^{- \frac{1}{2}} \epsilon P I P \abs{ \mathi \epsilon }^{-\frac{1}{2}} \\
&= J \abs{ \mathi \epsilon }^{- \frac{1}{2}} \left( 2 G - \mathi \epsilon \right) \abs{ \mathi \epsilon }^{-\frac{1}{2}} = 2 J \abs{ \mathi \epsilon }^{- \frac{1}{2}} G \abs{ \mathi \epsilon }^{-\frac{1}{2}} - \1 \eqend{,}
\end{splitequation}
which is a symmetric and densely defined operator in $\hat{\mathcal{H}}$, and where we used the relation~\eqref{eq:relation_g_i}. By uniqueness of the polar decomposition and using that $\epsilon P I P \geq 0$, we see that the polar decomposition of $E$ reads $E = J \abs{E}$ with $\abs{E} = \abs{ \mathi \epsilon }^{- \frac{1}{2}} \epsilon P I P \abs{ \mathi \epsilon }^{-\frac{1}{2}}$.

In terms of $E$, the modular Hamiltonian $K$~\eqref{eq:ilndelta_resolvent_gi} on $\hat{\mathcal{H}}$ reads
\begin{splitequation}
\label{eq:ilndelta_subspace_e}
K &= 2 \abs{ \mathi \epsilon }^{-\frac{1}{2}} \int_1^\infty \frac{E}{\1 - t^2 E^2} \total t \, \abs{ \mathi \epsilon }^\frac{1}{2} \\
&= 2 \abs{ \mathi \epsilon }^{-\frac{1}{2}} J \int_1^\infty \frac{\abs{E}}{\1 - t^2 \abs{E}^2} \total t \, \abs{ \mathi \epsilon }^\frac{1}{2} \\
&= - 2 \abs{ \mathi \epsilon }^{-\frac{1}{2}} J \arcoth(\abs{E}) \, \abs{ \mathi \epsilon }^\frac{1}{2} \eqend{,}
\end{splitequation}
where we again employed the integral representation~\eqref{eq:arccoth_integral} of the $\arcoth$. To ensure that this is a sensible result, we must have $\abs{E} \geq \1$, which again follows from positivity of $G$. Namely, the result~\eqref{eq:epspip_ieps_positive} still holds, and using that $\mathi \epsilon = J \abs{\mathi \epsilon}$ with $J^2 = \1$ we conclude that
\begin{equation}
\epsilon P I P - \abs{\mathi \epsilon} \geq 0
\end{equation}
on the intersection of their domains of definition, and thus
\begin{equation}
\abs{E} = \abs{ \mathi \epsilon }^{- \frac{1}{2}} \left( \epsilon P I P - \abs{\mathi \epsilon} \right) \abs{ \mathi \epsilon }^{-\frac{1}{2}} + \1 \geq \1 \eqend{.}
\end{equation}

The result~\eqref{eq:ilndelta_subspace_e} is again very similar to the result of Arias et al.~\cite{ariascasinihuertapontello2018}, who give the modular Hamiltonian as a second-quantized operator on Fock space. However, they also work with a discretized theory, where all operators are bounded. We then define
\begin{equation}
V = \frac{1}{2} \abs{ \mathi \epsilon }^{-\frac{1}{2}} E \abs{ \mathi \epsilon }^\frac{1}{2} = ( \mathi \epsilon )^{-1} G - \frac{1}{2} \eqend{,}
\end{equation}
which is weakly positive in the sense of Wigner, and hence has the same spectrum as $\frac{1}{2} E$. As in the scalar field case, $V$ thus has a spectral decomposition of the form~\cite{wigner1963}
\begin{equation}
V = \frac{1}{2} \int_0^M \lambda \total \left( \abs{ \mathi \epsilon }^{-\frac{1}{2}} E_\lambda \abs{ \mathi \epsilon }^\frac{1}{2} \right) \eqend{,}
\end{equation}
where $E_\lambda$ are the self-adjoint projections in the spectral decomposition of $E$ and $M = \norm{ E }$. This allows us to define functions of $V$ by spectral calculus, and we can write our result~\eqref{eq:ilndelta_subspace_e} as
\begin{equation}
\label{eq:lndelta_v_bounded}
K = - 2 \arcoth(2 V) = \ln\left( \frac{2 V - \1}{2 V + \1} \right) \eqend{,}
\end{equation}
which coincides with the result of Arias et al.~\cite{ariascasinihuertapontello2018}, except for a factor of $\abs{ \mathi \epsilon }^{-1}$ multiplying the expression~\eqref{eq:lndelta_v_bounded} on the right.

\subsection{KMS condition}
\label{sec:kms}

Lastly, we would like to show that the expression~\eqref{eq:ilndelta_resolvent_gi} for the modular Hamiltonian also follows from the KMS condition. This condition reads as follows: Given two elements $a,b$ in the von Neumann algebra, the function $F(t) = \omega\left( \sigma_t(a) b \right)$ admits an analytic continuation to the strip $- 1 < \Im t < 0$ with a continuous extension to the boundary $\Im t = -1$, and its boundary value there is given by $F(t-\mathi) = \omega\left( b \sigma_t(a) \right)$. Here, $\sigma_t(a) = \Delta^{\mathi t} a \Delta^{- \mathi t}$ is the modular flow generated by the modular Hamiltonian $\ln \Delta$, and $\omega$ is the state on the algebra that enters the construction. In our case, the von Neumann algebra is the algebra of Weyl operators with support inside the region $R \subset \Sigma$, and the state is the quasifree state on Fock space whose restriction to the one-particle level has the two-point function $G$.

Since the modular Hamiltonian on Fock space is the second quantization of the one-particle modular Hamiltonian~\cite{figlioliniguido1989,figlioliniguido1994}, for a Weyl operator $W(f)$ in the Fock representation we have the modular flow $\Delta^{\mathi t} W(f) \Delta^{- \mathi t} = W(f^t)$, where $f^t$ describes the flow at the one-particle level. Concretely, for a vector $f \in S_{R,\mathbb{R}}$ with support inside the region $R \subset \Sigma$, we have
\begin{equation}
f^t_a(x) = \sum_{b=1}^2 \int_R L_{ab}(t,x,y) f_b(y) \total y \eqend{,}
\end{equation}
where $L$ is an integral kernel that is related to the integral kernel of $K$~\eqref{eq:ilndelta_resolvent_gi} according to
\begin{equation}
\label{eq:relation_k_lndelta}
\mathi K_{ab}(x,y) = \left[ \frac{\partial}{\partial t} L_{ab}(t,x,y) \right]_{t = 0} \eqend{.}
\end{equation}
From the group property $\sigma_t( \sigma_s(a) ) = \sigma_{s+t}(a)$ of the modular flow, we also obtain the condition
\begin{equation}
\label{eq:k_semigroup}
L_{ab}(t+s,x,y) = \sum_{c=1}^2 \int_R L_{ac}(t,x,z) L_{cb}(s,z,y) \total z \eqend{,}
\end{equation}
which together with the relation~\eqref{eq:relation_k_lndelta} uniquely determines $L$. If we take $a = W(f)$ and $b = W(g)$ and use that
\begin{equation}
\omega\left( W(f) W(g) \right) = \mathe^{- \mathi \sigma(f,g)} \mathe^{- \frac{1}{2} \mu(f+g,f+g)} = \mathe^{- \frac{1}{2} G(f,f) - \frac{1}{2} G(g,g) - G(f,g)} \eqend{,}
\end{equation}
the KMS condition results in
\begin{equation}
G(f^{t - \mathi},f^{t - \mathi}) + 2 G(f^{t - \mathi},g) = G(f^t,f^t) + 2 G(g,f^t) \eqend{.}
\end{equation}
Consider the terms linear in $g$, for which this reduces to
\begin{splitequation}
&\sum_{a,b,c=1}^2 \iiint_R G_{ab}(x,y) L_{ac}(t-\mathi,x,z) f_c(z) g_b(y) \total x \total y \total z \\
&= \sum_{a,b,c=1}^2 \iiint_R G_{ab}(x,y) g_a(x) L_{bc}(t,y,z) f_c(z) \total x \total y \total z \eqend{,}
\end{splitequation}
and since $f$ and $g$ are arbitrary we obtain the relation
\begin{equation}
\label{eq:g_k_relation_kms}
\sum_{b=1}^2 \int_R G_{ba}(y,x) L_{bc}(t-\mathi,y,z) \total y = \sum_{b=1}^2 \int_R G_{ab}(x,y) L_{bc}(t,y,z) \total y
\end{equation}
that connects the integral kernel $G$ of the two-point function restricted to $R$ with $L$.

Interpreting $L$ as a convolution operator on $\hat{\mathcal{H}}$, the relation~\eqref{eq:g_k_relation_kms} reads $G^\text{T} L(t-\mathi) = G L(t)$, and the condition~\eqref{eq:k_semigroup} translates into $L(s+t) = L(t) L(s)$. It is easy to see that the unique solution of this system reads
\begin{equation}
\label{eq:kms_kl}
L(t) = \mathe^{\mathi t K} \quad\text{with}\quad K = - \ln\left[ G^{-1} G^\text{T} \right] \eqend{.}
\end{equation}
From the decomposition~\eqref{eq:g_restricted}, we also obtain
\begin{equation}
G_{ba}(y,x) = G_{ab}(x,y) - \mathi \epsilon_{ab} \delta(x,y)
\end{equation}
where we used that $X$ and $P$ are symmetric operators, or in operator notation $G^\text{T} = G - \mathi \epsilon$, such that the result~\eqref{eq:kms_kl} for $K$ can be written as
\begin{equation}
\label{eq:kms_k_g_relation}
K = - \ln\left[ \1 - G^{-1} \mathi \epsilon \right] = - \ln\left[ \1 - \left[ ( \mathi \epsilon )^{-1} G \right]^{-1} \right] \eqend{.}
\end{equation}
On the other hand, changing the integration variable in~\eqref{eq:ilndelta_resolvent_gi} as $t \to 1/t$ and performing a partial fraction decomposition (using the second resolvent identity), we obtain
\begin{splitequation}
K &= 2 \int_0^1 \frac{2 ( \mathi \epsilon )^{-1} G - \1}{t^2 - \left[ 2 ( \mathi \epsilon )^{-1} G - \1 \right]^2} \total t \\
&= \int_0^1 \left( \frac{1}{t - 2 ( \mathi \epsilon )^{-1} G + \1} - \frac{1}{t + 2 ( \mathi \epsilon )^{-1} G - \1} \right) \total t \\
&= - \int_0^1 \frac{1}{t + ( \mathi \epsilon )^{-1} G - \1} \total t = - \ln\left( \1 - \frac{1}{( \mathi \epsilon )^{-1} G} \right) \eqend{,}
\end{splitequation}
where we performed the additional variable changes $t \to 1-2t$ in the first integral and $t \to 2t-1$ in the second one. We have thus full agreement with the relation~\eqref{eq:kms_k_g_relation} derived from the KMS condition. The connection between the KMS condition and the modular flow was previously derived for chiral conformal fields, in particular the $\mathrm{U}(1)$ current~\cite{hollands2021}, but we see that it also holds for non-conformal free scalar fields.

\appendix

\begin{acknowledgments}
This work has been funded by the Deutsche Forschungsgemeinschaft (DFG, German Research Foundation) --- project no. 396692871 within the Emmy Noether grant CA1850/1-1.
It is a pleasure to thank Horacio Casini, Igor Khavkine and Christoph Minz for discussions, and Leonardo Sangaletti for discussions and a critical reading of the manuscript.
\end{acknowledgments}

\appendix

\bibliography{references}
\addcontentsline{toc}{section}{References}

\end{document}